\documentclass[twoside]{dis04}
\usepackage{
color}
\def\runauthor{D.\,Yu.\,Ivanov et al}
\def\shorttitle{Exclusive production of vector mesons in QCD}
\begin{document}

\title{Exclusive production of vector mesons in QCD
}

\author{D.\,Yu.\,Ivanov$^{+}$, G.\, Krasnikov$^\dag$ and L.\, Szymanowski$^*$}

\address{$^{+}$Sobolev Institute of Mathematics RAS,
 630090 Novosibirsk, Russia\\
$^\dag$Department of Theoretical Physics, St.Petersburg State University, \\
198904, St. Petersburg, Russia \\
$^*$So{\l}tan Institute for Nuclear Studies, Hoza 69, 00-681 Warsaw, Poland\\
E-mail: lechszym@fuw.edu.pl }

\maketitle

\abstracts{We compare results of our studies \cite{rho,heavy} on
 vector meson production
 within the QCD factorization at
next-to-leading order with HERA data.
}

\section{Introduction}

\vspace*{-0.3cm}The exclusive vector meson $V$ electro- and photo-productions in
photon-proton collisions, $\gamma\; p \to V\; p$, are the subject of HERA
collider experiments. The primary motivation for the interests in
these processes, both for the light $\rho-$meson electroproduction
and for the
heavy $J/\Psi-$ or eventually $\Upsilon-$meson photoproduction,
is that they can constrain
 gluon density in a proton. To achieve this aim it is necessary to
minimalize the theoretical uncertainties involved in determination of the
coefficient functions (CF), i.e.
the perturbatively calculable hard part of the scattering amplitudes.
The first step in this direction consists in going beyond the leading Born
approximation (LO) and requires
calculation of the scattering amplitudes at the next-to-leading order (NLO).
Recently such studies were completed both, for the processes with light
$\rho-$meson \cite{rho} and for those with heavy vector meson \cite{heavy}.
The theoretical framework of these researches  was the QCD factorization
which permits to write the scattering amplitude in a form of the
 convolution
(in fractions of longitudinal momenta) of the perturbatively calculable
CF with  nonperturbative distribution amplitude (DA)
of produced meson $V$ and the generalized parton distributions (GPDs)
in a target proton.

We refer the reader to Refs. \cite{rho,heavy} for
review of the literature on the subject,
for details of derivation and complete set of results. Below we present
a first comparison of these new results with  HERA data.

Let us remind  the main source of
theoretical uncertainties of our approach. The QCD factorization
involves the dependence of the scattering amplitude on two, a priori
independent, scales: the factorization scale $\mu_F$ and the
renormalization scale $\mu_R$. The dependence of the LO scattering amplitudes
on these scales is usually quite strong. By taking into account
NLO terms of the amplitudes one hopes to obtain a more consistent
description in which the
dependence of the amplitudes on $\mu_F$ and
$\mu_R$ is smoother than in the LO case
 and that the NLO corrections are small in comparison with LO terms. 
\vspace*{-0.5cm}

\section{Electroproduction of $\rho-$meson}

\vspace*{-0.2cm}Electroproduction of longitudinally polarized
neutral $\rho-$meson from the longitudinally polarized photon
on the proton target,
$\gamma_L^*(Q^2) \,p \to \rho_L \,p$, is a classical
 process  studied within the QCD factorization approach.
The hard scale required for the validity  of perturbative analysis
is supplied by the photon virtuality $Q^2$, the $\gamma^*_L \to \rho_L$
transition dominates.

Fig.~1 shows the comparison of theoretical predictions
of Ref.~\cite{rho} for
$\sigma(Q^2)$ with HERA data  \cite{Kreisel}.
The curves denoted by M  and C are obtained with two different input GPDs
based on forward parton distributions: MRST2001 and CTEQ6M, respectively
\cite{Freund}.
The factorization scale $\mu_F$ is assumed to be equal to
 the kinematical
hard scale $Q$ of the process.
On the other hand, the renormalization scale $\mu_R$
is
fixed in two different ways: for solid curves $\mu_R=\mu_F$, whereas for
dashed curves $\mu_R=Q/\sqrt{e}$. The last condition corresponds to the BLM
(Brodsky-Lepage-McKenzie) prescription. Our general conclusion is
that the account of the NLO terms
results in a  better  qualitative agreement of predictions with the data.
On the other hand,
the comparison of different curves shows a
strong dependence of theoretical predictions on
the input GPDs, especially   for $Q^2$ smaller
than 10~GeV$^2$.

\begin{figure}[h]
\label{1}
\vspace*{-0.4cm}
\begin{center}
\centerline{\epsfxsize=2.4in\epsfbox{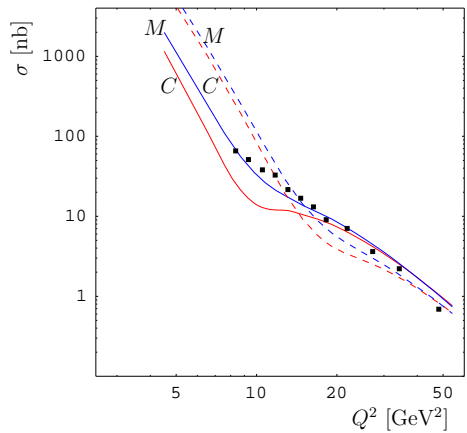}}
\caption[*]{ $\sigma(Q^2,W=95\mbox{GeV})$ as a function
of $Q^2$ for two NLO GPDs:
MRST2001 (denoted by M) and CTEQ6M (denoted by C).
The factorization scale
$\mu_f$ is assumed to be equal to $Q$. For the solid lines
 $\mu_R=\mu_F$, for the dashed lines
$\mu_R=Q/\sqrt{e}$.
The data points are taken from  \cite{Kreisel}.
}
\vspace*{-1cm}
\end{center}
\end{figure}

In Fig.~2 we show the dependence of $\sigma(W)$ on the
$\gamma^*-p$ cms scattering energy $W$. Each plot corresponds
to a different
value of virtuality $Q^2$ reported in \cite{Kreisel} with the
unchanged labeling of curves. We
observe again the strong behaviour of predictions on input GPDs.
Additionally, our predictions depend on assumed relations between
factorization and renormalization scales.
This last aspect is analyzed in more details in Fig.~3. It shows
a dependence of predictions for $d\sigma/dt|_{t=0}$
on the scale $\mu_F$ compared with three different
data points measured by ZEUS; on each plot they are
denoted by a horizontal line. We use also, the same as
previously prescriptions for fixing $\mu_R$.
Closer look at Fig.~3 shows that the inclusion
of NLO terms in the analysis results in a weaker
dependence of predictions on $\mu_F$ than
in the case when only the LO terms are kept. Nevertheless it remains still
quite strong which indicates that the NLO corrections are big in
comparison with LO contribution.

\begin{figure}[h]
\label{2}
\vspace*{-0.4cm}
\begin{center}
\centerline{\epsfxsize=5in\epsfbox{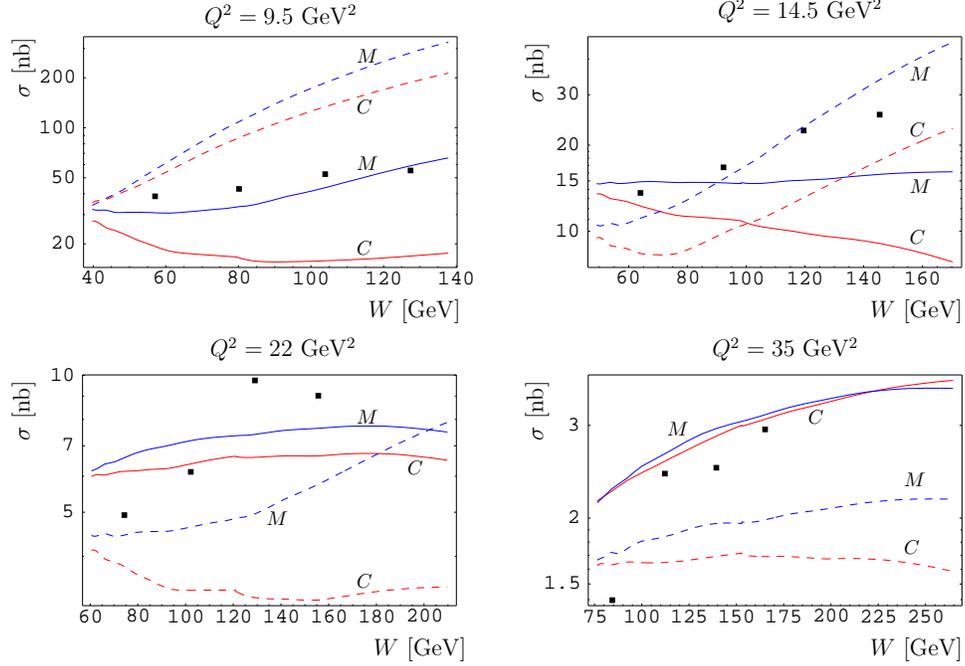}}
\caption[*]{ The cross section $\sigma(W)$ as
a function of $W$ for four
different values of $Q^2$ and for two NLO GPDs:
MRST2001 (denoted by M) and CTEQ6M (denoted by C). The factorization scale
$\mu_f$ is assumed to be equal to $Q$. The solid lines correspond to the
$\mu_R=\mu_F$, for the dashed lines $\mu_R=Q/\sqrt{e}$.
The data points are taken from \cite{Kreisel}    }
\vspace*{-1cm}
\end{center}
\end{figure}

This feature is illustrated in Fig.~4 which shows
the relative magnitude of different contributions to the 
scattering amplitude. Firstly we note that, as expected in the small$-x$
HERA kinematics, the imaginary part of  scattering amplitude is
much larger than its real part.
The comparison of LO gluon (gB) and quark (qB)
contributions with those including LO and NLO terms (curves denoted as full)
shows that the NLO corrections are large and mostly have opposite signs
than the Born contributions. Consequently the final predictions
are the result of strong 
cancellations between the LO and the NLO contributions.

\begin{figure}[h]
\label{3}
\begin{center}
\centerline{\epsfxsize=5in\epsfbox{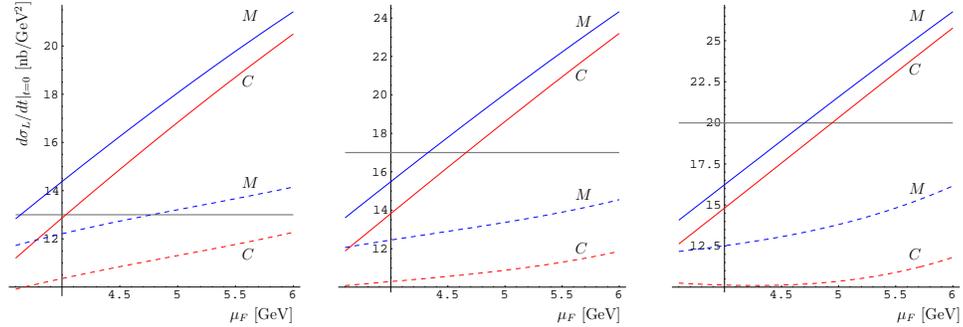}}
\caption[*]{
$d\sigma_L/dt (t=0)$ as a function of $\mu_F$ compared with three experimental
points $13,\;17,\;20\,$nb/GeV$^2$ (horizontal lines from left to
right) corresponding to three values of
$W=\,80,\;110,\;150\,$GeV with $Q^2=27\,$GeV$^2$.
The labeling of curves is  the same as in
Figs. 1 and 2. The data points are taken from \cite{BreitwegRHO}
}
\vspace*{-1cm}
\end{center}
\end{figure}

\begin{figure}[h]
\label{4}
\begin{center}
\centerline{\epsfxsize=5in\epsfbox{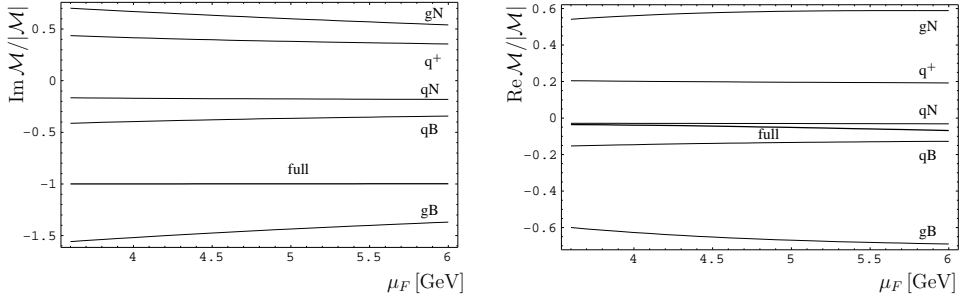}}
\caption[*]{For  $Q^2=27\,$GeV$^2$, $W=110\,$GeV different 
contributions  to
$Im\, {\cal M} /|{\cal M}|$ and $Re\, {\cal M}/ |{\cal M}|$
in dependence on $\mu_F$
(for $\mu_R=\mu_F$ and CTEQ6M):
$gB$ -- gluon contr. at LO, $qB$ -- quark contr.  at LO,
$gN$ -- gluon contr. at NLO, $qN$ and $q^+$ -- quark contr. 
 at NLO  (see \cite{rho} for more details),
full -- all terms LO and NLO included. $|{\cal M}|$ contains always
LO plus NLO terms.
 }
\vspace*{-1cm}
\end{center}
\end{figure}

\section{Photoproduction of heavy vector mesons}

In Ref.~\cite{heavy} we studied
 a photoproduction of
transversely polarized vector mesons $V=\Upsilon$, $J/\Psi$,
on a proton, $\gamma_T\;p\to V_T \;p$. The hard scale in this process
is supplied by the mass of heavy meson.

As was already mentioned,  the inclusion of NLO terms leads to
a weaker dependence of
predictions on scales  $\mu_F$ and $\mu_R$. In the case of
$\Upsilon-$meson photoproduction this fact is illustrated in Fig.~5.
The left plot shows the comparison of ZEUS and H1 data with LO predictions
(when only gluon GPD contributes)
by assuming that $\mu_F=\mu_R$ and they vary in the interval
$[1.3-7]\,$GeV. In
the right plot the same comparison is done with NLO terms included.
As result it leads to
a weaker variation of theoretical predictions on values of these scales.

\begin{figure}[h]
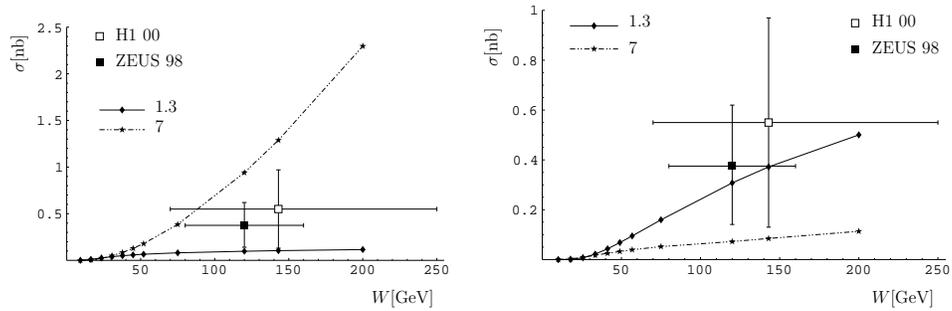

\label{5}
\begin{center}
\scalebox{0.55}{
\input{fig13.pstex_t}
}\,\,\,\,
\scalebox{0.55}{
\input{fig14.pstex_t}
}
\caption[*]{
The cross section of the $\Upsilon$ photoproduction;
theoretical predictions at LO (left figure) and NLO (right figure)
for the scales
$\mu_F=\mu_R=[1.3,7]\mbox{ GeV}$. 
The data are from ZEUS \cite{Breitweg:1998ki} and H1 \cite{Adloff:2000vm}.
}
\vspace*{-0.8cm}
\end{center}
\end{figure}

Fig.~6 is the $\Upsilon-$meson production analog of Fig.~4
from $\rho-$meson case. We observe similar feature 
that the imaginary part
of scattering amplitude with LO and
NLO terms included (denoted as total)
is larger than the corresponding real part. 
We observe also in this case that the NLO gluon and quark
corrections are large in comparison with LO (Born) contributions and
have opposite signs.
Thus again the final predictions are result of strong cancellations
between the LO and the NLO terms.

\begin{figure}[h]
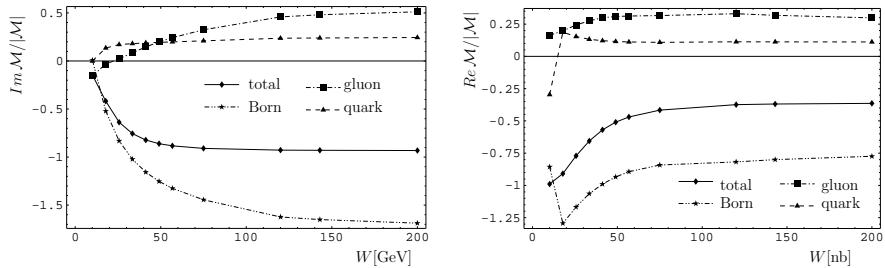

\label{6}
\begin{center}
\scalebox{0.5}{
\input{fig15c.pstex_t}
}\,\,\,\,
\scalebox{0.5}{
\input{fig15d.pstex_t}
}
\caption[*]{
$\Upsilon$ photoproduction, NLO prediction for $\mu_F=\mu_R =4.9\mbox{
GeV}$ and its decomposition into different contributions, see text.}
\vspace*{-0.8cm}
\end{center}
\end{figure}

The  presently available and cleanest experimentally process which could
serve for study of 
 gluon distribution in a nucleon  is the
$J/\Psi$ photoproduction. The analog of Figs.~5,~6 for this case is
shown in Fig.~7.
By inspecting the top-left plot one could at first sight
to conclude that
 inclusion on the NLO terms leads to predictions reflecting better
 the behaviour of data. But a closer look into the left-bottom plot shows
that the difference between LO and NLO terms are larger than
in the case of $\Upsilon$ production, in fact they have even
opposite signs. Moreover, as can be seen also in the right-top plot,
the LO plus NLO
imaginary part crosses zero at some point, which seems to suggest that
our predictions are not numerically stable. The obvious reason for that is
 too small value of
$J/\Psi-$meson mass as a hard scale.

\begin{figure}[h]
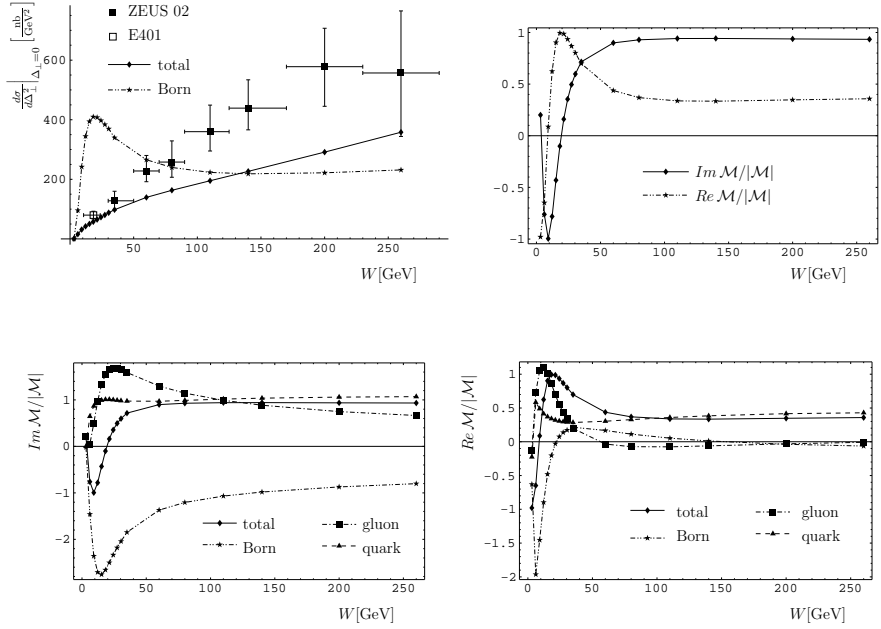

\begin{center}
\scalebox{0.5}{
\input{fig17a.pstex_t}
}\,\,\,\,\,\,
\scalebox{0.5}{
\input{fig17b.pstex_t}
}\\[1cm]
\hspace*{1mm}\scalebox{0.5}{
\input{fig17c.pstex_t}
}\,\,\,\,
\scalebox{0.5}{
\input{fig17d.pstex_t}
}
\caption[*]{
The differential  cross section for $J/\psi$ photoproduction,
NLO predictions for $\mu_F=\mu_R =1.52\mbox{ GeV}$,  and the data from
 E401
\cite{Binkley:1981kv} and
ZEUS \cite{Chekanov:2002xi}.
The labeling of the curves is the same as in Fig.~\ref{6}.}
\vspace*{-0.4cm}
\end{center}
\label{fig7}
\end{figure}

\section{Conclusions}

The general conclusion which comes out from the comparison of
the theoretical predictions on vector mesons production at NLO
with experiment is
a qualitative improvement of the description at NLO in comparison with
the one at LO.
This fact is of some value by itself since
 these results are to big extent model independent.
Nevertheless, at HERA kinematics the NLO contributions are large in
comparison with the LO terms which calls for the all-order
resumation of these terms. Another possibility, related to production
of heavy mesons, is to extend the analysis of Ref.~\cite{heavy}
to the electroproduction processes involving
virtual photon. This would lead to a larger values of hard scale
and hopefully to a better numerical stability of predictions.
Finally let us also mention that the non-perturbative power corrections
are probably not negligible at HERA kinematics. 

\section*{Acknowledgments} Work of L.Sz. is partially supported by
 the French-Polish scientific agreement Polonium. D.~I. is supported by 
grants DFG 436, RFBR 03-02-17734.

\end{document}